\documentclass[aps,prl,preprint,groupedaddress,showpacs]{revtex4}

\usepackage{graphicx}

\begin{document}


\title{A one-dimensional Vlasov-Maxwell equilibrium for the force-free Harris sheet}


\author{Michael G. Harrison}
\author{Thomas Neukirch}

\affiliation{School of Mathematics and Statistics, University of St. Andrews, St. Andrews, KY16 9SS, United Kingdom}


\date{\today}

\begin{abstract}
In this paper the first non-linear force-free Vlasov-Maxwell equilibrium is presented. One component of the equilibrium magnetic field has the same spatial structure as the Harris sheet, but whereas the Harris sheet is kept in force balance by pressure gradients, in the force-free solution presented here force balance is maintained by magnetic shear. Magnetic pressure, plasma pressure and plasma density are constant. The method used to find the equilibrium is based on the analogy of the one-dimensional Vlasov-Maxwell equilibrium problem to the motion of a pseudo-particle in a two-dimensional conservative potential. This potential is equivalent to one of the diagonal components of the plasma pressure tensor. After finding the appropriate functional form for this pressure tensor component, the corresponding distribution functions can be found using a Fourier transform method. The force-free solution can be generalized to a complete family of equilibria that describe the transition between the purely pressure-balanced Harris sheet to the force-free Harris sheet. 

\end{abstract}


\maketitle


Force-free magnetic fields, i.e. magnetic fields satisfying
\begin{eqnarray}
(\nabla \times \mathbf{B}) \times \mathbf{B} & = &\mathbf{0},
\label{forcefreeequation}\\
\nabla \cdot \mathbf{B} = 0, \label{divB0}
\end{eqnarray}
are important for modelling low-$\beta$ plasmas in laboratory, space and astrophysical applications\cite{Marshbook}.  Equation (\ref{forcefreeequation}) implies that $\nabla\times \mathbf{B}$ (basically the electric current density) has to be aligned with the magnetic field, i.e. $\nabla \times \mathbf{B} = \alpha \mathbf{B}$. The scalar function $\alpha$ is constant along magnetic field lines due to Eq. (\ref{divB0}), but can vary from field line to field line. If $\alpha$ does not vary from field line to field line, but is globally constant we get the case of linear force-free fields (sometimes also called constant-$\alpha$ fields). All other force-free fields are called non-linear force-free fields.
 
Using magnetohydrodynamics (MHD) many useful linear and non-linear force-free magnetic fields can be found analytically, especially if translational or rotational symmetry of the solutions is assumed (see e.g. \cite{Marshbook, Tassi-2008}). This is completely different if one considers collisionless Vlasov-Maxwell (VM) equilibria (see e.g. the discussion in \cite{Tassi-2008}). So far, only one-dimensional linear force-free VM equilibria have been found\cite{Sestero-1967,Channell-1976,Bobrova-1979,Correa-Restrepo-1993,Bobrova-2001}, and, to the best of our knowledge, no non-linear force-free VM equilibria are known.

One-dimensional (1D) VM equilibria are frequently used as a starting point for studies of waves and instabilities in collisionless plasmas. One of the most commonly used 1D VM equilibria is the 
Harris sheet\cite{Harris-1962}, with $\mathbf{B} (z) = B_0 \tanh(z/L) \mathbf{e}_x$ and 
$\mathbf{j} (z) = B_0/(\mu_0 L) \cosh^{-2} (z/L) \mathbf{e}_y$, so the current density is perpendicular to the magnetic field. The force balance is maintained by a pressure gradient. Often, a constant magnetic field in the $y$-direction (guide field) is added, which, if sufficiently strong, is used to mimic a force-free field. It is clear that through introducing a guide field the current density is partially field-aligned, but the strength of the guide field is completely decoupled from the strength of the current density. In force-free fields a stronger current density would lead to a stronger shear of the magnetic field as the two are closely coupled. Furthermore, a constant magnetic field will not add any free energy to the system, whereas one expects an increase in free energy if the magnetic shear in a force-free field is increased. As a final point we mention that force-free equilibria will have constant density and pressure, whereas the Harris sheet plus guide field has the same pressure and density gradients as the Harris sheet itself. This may be an important difference in studies of, for example, magnetic reconnection (see e.g. \cite{Daughton-2003, Daughton-2004,Ricci-2004,Ricci-2004b,Ricci-2005}).
Some investigations of the stability and dynamics of the known linear force-free 1D VM equilibria have been undertaken\cite{Bobrova-2001,Li-2003,Bowers-2007}, but it is to be expected that non-linear force free equilibria will have new and interesting properties.



Generally VM equilibria can only be found easily for cases with spatial symmetries, and to obtain analytical force-free solutions one has to investigate situations with invariance along two coordinate directions.
In this paper we consider the case of translationally invariant VM equilibria depending only on one spatial coordinate, here taken to be $z$.

We assume that the magnetic field has components $B_{x}$ and $B_{y}$. The magnetic field components are written in terms of a vector potential $\mathbf{A}=(A_x,A_y,0)$ where 
$B_x= - dA_y/dz$ and $B_y= dA_x/dz$.

We assume a plasma consisting of two particle species of equal, but opposite charge (electrons and ions/protons).
Due to the symmetries of the system the three obvious constants of motion for each particle species are the Hamiltonian or particle energy for each species $s$, 
$
  H_{s}=\frac{1}{2}m_s (v_x^2+v_y^2+v_z^2)+q_{s}\phi,
$ 
the canonical momentum in the $x$-direction, $p_{xs}=m_{s}v_{x}+q_{s}A_{x}$,
and the canonical momentum in the $y$-direction, $p_{ys}=m_{s}v_{y}+q_{s}A_{y}$.
Here $\phi$ is the electric potential and $m_{s}$ and  $q_{s}$ are the mass and charge of each particle species. 
All positive functions $f_s $ satisfying the appropriate conditions for existence of the velocity moments and depending only on the constants of motion, $f_{s}=f_{s}(H_{s},p_{xs},p_{ys})$
are solutions of the steady-state Vlasov equation.

One can show\cite{Mynick-1979a,Harrison-2009} that for a quasi-neutral plasma, Ampere's law can be written as 
\begin{eqnarray}
  \frac{d^{2}A_{x}}{dz^2}&=&-\mu_{0}\frac{\partial P_{zz}}{\partial A_{x}} ,\label{amperex2}
\\
  \frac{d^{2}A_{y}}{dz^2}&=&-\mu_{0}\frac{\partial P_{zz}}{\partial A_{y}}, \label{amperey2}
\end{eqnarray}
where $P_{zz}(A_x,A_y)$ is the $zz$-component of the plasma pressure tensor, defined by
\begin{equation}
P_{zz} = \sum_{s} \int_{-\infty}^{\infty} m_{s}v_{z}^2f_{s}d^3v .
\label{Pzz}
\end{equation}
Equations (\ref{amperex2}) and (\ref{amperey2}) can be immediately integrated once to give the
force balance condition across the sheet as 
\begin{equation}
\frac{B^2}{2 \mu_0} + P_{zz} = P_T=\mbox{constant}.
\label{totalpressure}
\end{equation}
Due to Eq. (\ref{forcefreeequation}) a force-free equilibrium satisfies the conditions
$B^2=$ constant and $P_{zz} =$ constant
separately.

The 1D VM equilibrium equations  (\ref{amperex2}) and (\ref{amperey2})  are equivalent to the equations of motion of a (pseudo-)particle in a conservative 2D pseudo-potential $P_{zz}(A_x, A_y)$\cite{Channell-1976, Mynick-1979a}.
The position of the pseudo-particle is given by $A_x$, $A_y$ with the pseudo-time given by $z$. The energy (Hamiltonian) of this pseudo-particle is given by the total pressure defined in Eq. (\ref{totalpressure}) (modulo a factor $\mu_0$) 
$E =  [(d A_x/dz)^2+  (d A_y/dz)^2 ]/2+
\mu_0 P_{zz}(A_x,A_y)$.
One can show that a force-free VM solution corresponds to a pseudo-particle trajectory that is identical to a contour of the pseudo-potential\cite{Harrison-2009}. This is easily possible for attractive central potentials which have circular contours and also allow circular pseudo-particle orbits. These circular orbits correspond to the known linear force-free solutions\cite{Sestero-1967,Channell-1976,Bobrova-1979,Correa-Restrepo-1993,Bobrova-2001}, which, as far as we are aware, are the only known force-free VM solutions. For finding nonlinear force-free solutions we obviously need to find a pseudo-potential ($P_{zz}(A_x,A_y)$) which is not a central potential, but still allows a solution to Eqs. (\ref{amperex2}) and (\ref{amperey2}) that is identical with an equipotential line.

Channell\cite{Channell-1976} showed how, by making a number of sensible assumptions, a transform method can be used to determine a class of distribution functions for a known $P_{zz}(A_x,A_y)$. Mynick et al.\cite{Mynick-1979a} generalized this method and used it to determine the distribution functions numerically. In this paper we will first determine a function $P_{zz}(A_x,A_y)$ for the force-free Harris sheet and then use Channell's method to find the corresponding distribution functions.

\begin{figure*}
\includegraphics[width=0.8\textwidth]{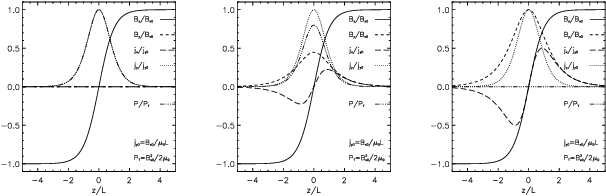}%
\caption{\label{figure1} The magnetic field, current density and pressure profiles as functions 
of $z/L$ for the Harris sheet (left panel), the force-free Harris sheet (right panel) and 
intermediate case (middle panel).}
\end{figure*}
The magnetic field of the force-free Harris sheet solution is given by 
(see also Fig. \ref{figure1})
\begin{eqnarray}
  B_x & = & B_0 \tanh(z/L), \\
  B_y & = & B_0\cosh^{-1} (z/L),
  \end{eqnarray}
 with $B_0$ the constant amplitude of the field and $L$ the sheet half width. Obviously we have $B_x^2+B_y^2 =  B_{0}^2$.
 One can easily see that $j_x =  B_0/(\mu_0 L) \tanh(z/L)/\cosh(z/L)$, $j_y = B_0/(\mu_0 L)\cosh^{-2}(z/L)$,
 giving $\alpha(z) = [L \cosh(z/L)]^{-1}$.
 
 The $x$-component of this magnetic field is identical to the Harris sheet $B_x$, but in this case the force balance is  maintained by the magnetic shear component $B_y$ instead of the plasma pressure.
The vector potential for the force-free Harris sheet field is found to be given by
 \begin{eqnarray}
  A_{x,ffh} & = & 2 B_0 L \arctan\left(\exp(z/L)\right), \label{axharrisff}\\
  A_{y,ffh} & = & -B_0 L \ln \left(\cosh \left(\frac{z}{L}\right)  \right), \label{ayharrisff}
\end{eqnarray}
in a convenient gauge.

In order to make analytical progress we assume that $P_{zz}$ has the form $P_{zz}(A_x,A_y) = P_1(A_x)+P_2(A_y)$.
The physical meaning of this assumption is that for each particle species there are two different particle populations that carry the components of the current density in the $x$- and the $y$-directions.
Eqs. (\ref{amperex2}) and (\ref{amperey2})
give the conditions 
 \begin{eqnarray}
  \left(\frac{d A_{x}}{dz}\right)^2 + 2 \mu_0 P_1(A_x) & = & 2\mu_0 P_{01}, \label{ffhpot1} \\
  \left(\frac{d A_{y}}{dz}\right)^2 + 2 \mu_0 P_2(A_y) & = & 2\mu_0P_{02}, \label{ffhpot2} 
\end{eqnarray}
where $P_{01}$ and $P_{02}$ are constants. Equations (\ref{ffhpot1}) and (\ref{ffhpot2}) will be used to find the appropriate $P_{zz}(A_x,A_y)$.

\begin{figure}
\begin{center}
\includegraphics[width=0.35\textwidth]{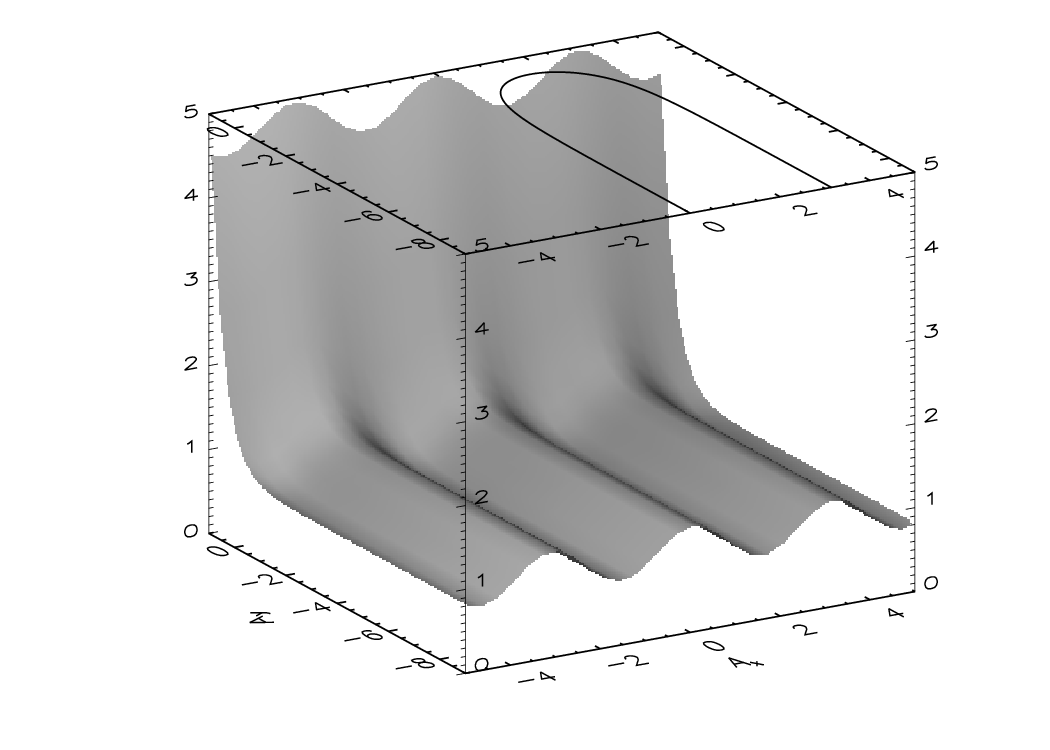}%
\end{center}
\caption{\label{figure2}  A surface plot of the pressure function $P_{zz}(A_x,A_y)$ for the force-free Harris sheet. The force-free Harris sheet solution is shown as a pseudo-particle trajectory at the top of the plot. It is identical with a contour of $P_{zz}$.}
\end{figure}

We substitute $A_{x,ffh}$ and $A_{y,ffh}$ into the first terms of Eqs. (\ref{ffhpot1}) and  (\ref{ffhpot2}) and then use that $\exp(z/L) = \tan(A_x/2B_0 L)$ and $\cosh(z/L)=\exp(-A_y/B_0 L)$ to obtain
\begin{equation}
P_{zz}= \frac{B_{0}^2}{2 \mu_0} \left[\frac{1}{2} \cos\left(\frac{2 A_x}{B_0 L}\right) + \exp\left(\frac{2 A_y}{B_0 L}\right) \right] + P_{03}.
\label{force-freepotential}
\end{equation}
A surface plot of 
$P_{zz}(A_x,A_y)$ is shown in Fig. \ref{figure2}. Above the surface plot the trajectory representing the force-free Harris sheet solution in the $A_x$-$A_y$-plane is shown. By construction it is identical to a contour of $P_{zz}(A_x,A_y)$. 

We use Channell's\cite{Channell-1976} Fourier transform method to solve the integral equation (\ref{Pzz}) for the distribution functions $f_s$. 
The method is based on the assumptions that  (a) the distribution functions have the form $f_s(H_s,P_{xs},p_{ys}) =f_{0s} \exp(-\beta_s H_s) g_s(p_{xs},p_{ys})$ and that  (b) the quasineutral electric potential $\phi_{qn}$ vanishes. The validity of the second assumption can be easily verified {\em a posteriori} and only requires the correct choice of parameters.
Applying the method we find that the required distribution functions are of the form 
\begin{eqnarray}
f_s &=& \frac{n_{0s}}{v_{th,s}^3} \exp(-\beta_s H_s) \left[\exp \left( \beta_s u_{ys} p_{ys} \right)
 \right. \nonumber \\
& &  \left.+ a_s \cos \left(\beta_s u_{xs}p_{xs} \right) 
   + b_s \right],
\label{nonlinearffreedist}
\end{eqnarray}  
where $v_{th,s} = (m_s\beta_s)^{-1/2}$ is the thermal velocity of particle species $s$ and $u_{xs}$, 
$u_{ys}$, $a_s$ and $b_s$ are constants with $0<a_s < b_s$. We have here reverted to the usual microscopic notation for the distribution functions. We will make the connection to the notation used previously by calculating $P_{zz}$ directly from the distribution function and then comparing the result with the expression (\ref{force-freepotential}). This is useful to relate the macroscopic quantities $B_0$ and $L$ to the microscopic parameters of the distribution function.
The first part of this distribution function is identical with the Harris sheet\cite{Harris-1962} distribution function, whereas the second part corresponds to a different particle population which carries the current in the $x$-direction and is responsible for the shear field $B_y(z)$.
When calculating  $P_{zz}$ from the distribution function one finds that it has the general structure
$P_{zz}=\sum_s \beta_s^{-1} \exp(-q_s\beta_s \phi) N_s(A_x,A_y)$,
where
\begin{eqnarray*}
N_s(A_x,A_y) &= & \sqrt{8 \pi^3} n_{0s}  \exp(\beta_s m_s u_{ys}^2/2) \nonumber \\
& & \mbox{\hspace{-1.6cm}}\times  [ \exp(\beta_s u_{ys} q_s A_y)  \nonumber \\
& & \mbox{\hspace{-1.4cm}} + a_s\exp(-\beta_s m_s (u_{xs}^2+u_{ys}^2)/2)\cos(\beta_s u_{xs} q_s A_x) \nonumber \\
& & \mbox{\hspace{-1.4cm}}+ 
b_s \exp(-\beta_s m_s u_{ys}^2/2)].
\end{eqnarray*}
The quasi-neutrality condition leads to $\phi_{qn} = [e(\beta_e+\beta_i)]^{-1}\ln(N_i/N_e)$.
The condition of vanishing quasi-neutral electric potential implies that $N_i (A_x,A_y)=N_e(A_x,A_y)$, which is true if
\begin{eqnarray*}
& &\mbox{\hspace{-0.3cm}}n_{0e} \exp(\beta_e m_e u_{ye}^2/2) =  n_{0i} \exp(\beta_i m_i u_{yi}^2/2) =n_0/\sqrt{8\pi^3} ,\\
& & \mbox{\hspace{-0.3cm}}a_e\exp(-\beta_e m_e (u_{xe}^2 + u_{ye}^2)/2 )=  \\
&& \mbox{\hspace{1.5cm}}a_i\exp(-\beta_i m_i (u_{xi}^2 + u_{yi}^2) /2) =a,\\
& &\mbox{\hspace{-0.3cm}}b_e\exp(-\beta_e m_e u_{ye}^2/2 )= b_i\exp(-\beta_i m_i  u_{yi}^2/2) =b,\\
%
&&-\beta_e u_{xe}  =  \beta_i u_{xi}, \\
&& -\beta_e u_{ye}  =  \beta_i u_{yi} .
\end{eqnarray*}
Supposing that $\beta_e$ and $\beta_i$ are given we have ten other parameters needing to satisfy only five equations, which is always possible.
 This provides the necessary {\em a posteriori} justification for using Channell's method. Using the notation often used for the Harris sheet (e.g. \cite{Schindlerbook})  $P_{zz}$  becomes
 \begin{eqnarray}
 & & \mbox{\hspace{-1.5cm}}P_{zz} =   \left(\frac{1}{\beta_e}+\frac{1}{\beta_i}\right) n_0 
  [\exp(-e\beta_e u_{ye} A_y)\nonumber \\
  & & \mbox{\hspace{1.5cm}} +a \cos(e\beta_e u_{xe} A_x) +b].
 \label{finalPzz}
 \end{eqnarray}
 
Comparison with Eq. (\ref{force-freepotential}) shows that for the force-free Harris sheet the connection between the microscopic notation and the original notation is given by
\begin{eqnarray}
\frac{B_0^2}{2 \mu_0} &=& \left(\frac{1}{\beta_e}+\frac{1}{\beta_i}\right) n_0 , \label{n0def} \\
L& = &\left(\frac{2\beta_i}{\mu_0e^2 n_0 u_{ye}^2 \beta_e(\beta_e +\beta_i)}\right)^{1/2},
\label{Ldef}\\
a &=& \frac{1}{2}  ,\label{adef} \\
b &=&  2 \mu_0P_{03}/B_0^2. \label{bdef}
\end{eqnarray}
Eq. (\ref{Ldef}) is especially important as it provides a relation between the length scale $L$ and the parameters of the distribution function. This, for example, directly links
$\alpha(z)=[L \cosh(z/L)]^{-1}$ derived from the general form of the magnetic field and current density to the microscopic parameters of the distribution function.






It is straightforward to see that for different parameter values the distribution function (\ref{nonlinearffreedist}) gives the complete family of equilibria describing the transition between the Harris sheet and the force-free Harris sheet. 
The intermediate cases have, written as functions of $z$,
\begin{eqnarray}
  B_x & = & B_{x0} \tanh(z/L) ,\\
  B_y & = & B_{y0}\cosh^{-1}(z/L) ,\\
  P_{zz} &=& P_0\cosh^{-2}(z/L) + P_{00},
  \end{eqnarray}
  where $P_0 + B_{y0}^2/2\mu_0= B_{x0}^2/2 \mu_0$.
Taking the limit $B_{y0}\to 0 $ gives the Harris sheet \cite{Harris-1962} and taking the limit $P_0 \rightarrow 0$ gives the force-free Harris sheet.

An appropriate $P_{zz}(A_x,A_y)$ can be determined in the same way as for the force-free Harris sheet and takes the form
\begin{equation}
 P_{zz}=  \frac{B_{x0}^2}{2 \mu_0}\exp\left(\frac{2 A_y}{B_{x0} L}\right)  
+ \frac{1}{2} \frac{B_{y0}^2}{2 \mu_0} \cos\left(\frac{2 A_x}{B_{y0} L}\right)  + P_{03}.
\label{combinedpotential}
\end{equation}
In this case a comparison between Eqs. (\ref{combinedpotential}) and (\ref{finalPzz}) shows that Eqs.
(\ref{n0def}),  (\ref{Ldef}) and (\ref{bdef}) do not change apart from $B_0$ becoming $B_{x0}$, but that we now also have
\begin{eqnarray}
B_{y0}& = &\left(\frac{2\mu_0 (\beta_e + \beta_i) n_0 u_{ye}^2}{\beta_e\beta_i u_{xe}^2}\right)^{-1/2},\\
a  &=& \frac{1}{2}  \frac{B_{y0}^2}{B_{x0}^2}.
\end{eqnarray}


As shown by Harrison and Neukirch\cite{Harrison-2009} one can deduce 
from one $P_{zz}(A_x,A_y)$  allowing a force-free VM solution an infinite number of other functions
$\bar{P}_{zz}(A_x,A_y)$ 
allowing the {\em same} force-free solution by using positive definite
 functions of the known $P_{zz}(A_x,A_y)$. We mention in particular the possibility of using an exponential function of the $P_{zz}$  presented here, as it would give rise to a product form for $P_{zz}$ instead of a sum. The distribution functions would also consist of products of functions of $p_{xs}$ and $p_{ys}$ instead of a sum. It is, however, unclear whether the method 
used in this paper would still allow for an analytical calculation of these distribution functions. 
 

This new family of VM equilibria will generate new possibilities for studies of linear and nonlinear instabilities of force-free current sheets. The stability of the VM equilibria presented here have yet to be investigated. We point out that the $p_{xs}$-dependent part of the distribution function may have multiple peaks in the $v_x$-direction and we suspect that this will give rise to instabilities. We also remark that although the $B_x(z)$ and $j_y(z)$-profiles are identical to the Harris sheet, $j_x(z)$ is antisymmetric with respect to $z=0$. This is closely linked to the fact that in the Harris sheet solution the spatial structure of the current density is determined by the density structure with the average velocity of each particle species being constant, whereas in the force-free solution presented here the particle density is constant and the spatial structure of the current density is determined by the spatial structure of the average velocity. Further investigations will be needed to clarify exactly what the implications are for the stability of the new solution, but on the basis of the physical differences just mentioned one would expect the stability properties of the force-free solution to differ considerably from those of the Harris sheet.
Apart from studying the stability properties of the solution class presented here, it will be also be very  interesting to investigate whether the general method employed here can be used to find other non-linear force-free solutions. 
\begin{acknowledgments}
The authors thank the referees for useful remarks and acknowledge financial support by the UK's Science and Technology Facilities Council and by the European Commission through the SOLAIRE Network (MTRN-CT-2006-035484).
\end{acknowledgments}


\end{document}